\documentclass[twocolumn,nofootinbib,amsmath,amssymb,a4paper]{revtex4}

\usepackage{graphicx}
\usepackage{epsfig}
\usepackage{dcolumn}
\usepackage{bm}

\begin{document}

\title{$V_{us}$ From Hadronic $\tau$ Decays}

\author{K. Maltman}
 \email{kmaltman@yorku.ca}
\affiliation{%
Dept. Mathematics and Statistics., York University, Toronto, ON CANADA 
M3J 1P3 and\\ CSSM, Univ. of Adelaide, Adelaide SA 5005 AUSTRALIA}%

\author{C.E. Wolfe}
 \email{wolfe@yorku.ca}
\affiliation{%
Dept. Physics and Astronomy, York University, Toronto, ON CANADA M3J 1P3}%

\begin{abstract}
We describe how complications encountered in the extraction of
$V_{us}$ from flavor-breaking hadronic $\tau$ decay sum rules, in 
particular those associated with the slower than previously anticipated
convergence of the relevant $D=2$ OPE series, can be successfully 
dealt with through judicious choices of sum rule weight. 
Problems with the previously proposed ``$(0,0)$ spectral weight'' version of 
the analysis, and the much improved situation produced by the use of a number 
of alternate, non-spectral-weights, are demonstrated. The
method is shown to promise a sub-$0.0010$-accuracy determination of $V_{us}$ 
once analysis of the improved strange spectral data 
from BABAR and BELLE is completed. Connections to the
$\tau$ versus electroproduction puzzle in the evaluation
of the leading order hadronic contribution to $(g-2)_\mu$ are also discussed.
\end{abstract}

\maketitle

\section{Background/Terminology}
The ratios $\Gamma [\tau^- \rightarrow \nu_\tau
\, {\rm hadrons}_{V/A;ij}\, (\gamma )]/
\Gamma [\tau^- \rightarrow \nu_\tau e^- {\bar \nu}_e (\gamma)]$
($\equiv\, R_{V/A;ij}$) involving inclusive flavor $ij=ud,us$, 
vector (V) or axial vector (A) current induced
hadronic $\tau$ decay widths, are related to the spectral
functions, $\rho^{(J)}_{V/A;ij}$, of the spin $J=0$ and $1$
parts, $\Pi^{(J)}_{V/A;ij}$, of the corresponding current-current correlators
by~\cite{bnpetc}
\begin{eqnarray}
&&{\frac{R_{V/A;ij}}{\left[ 12\pi^2\vert V_{ij}\vert^2 S_{EW}\right]}}
=\int^{1}_0\, dy_\tau \,
\left( 1-y_\tau\right)^2 \nonumber\\
&&\quad\left[ \left( 1 + 2y_\tau\right)
\rho_{V/A;ij}^{(0+1)}(s) - 2y_\tau \rho_{V/A;ij}^{(0)}(s) \right]
\label{basictaudecay}
\end{eqnarray}
where $y_\tau =s/m_\tau^2$, $V_{ij}$ is the
flavor $ij$ CKM matrix element, and $S_{EW}$ is
a short-distance electroweak correction.
Analogous spectral integrals, $R_{V/A;ij}^{(k,m)}$, are
obtained from $R_{V/A;ij}$ by rescaling the integrand with
$(1-y_\tau )^ky_\tau^m$ before integration. In what follows we
denote by $R^w_{ij}(s_0)$ the weighted spectral integral
from threshold to $s=s_0\leq m_\tau^2$ involving the analytic weight, $w(s)$,
and either of the correlators $\Pi^{(0+1)}_{V/A;ij}(s)$ or 
$s\Pi^{(0)}_{V/A;ij}(s)$. All spectral integrals discussed above
satisfy the general finite energy sum rule (FESR) relation
\begin{equation}
\int_0^{s_0}ds\, w(s) \rho (s) =\, {\frac{-1}{2\pi i}}\,
\oint_{\vert s\vert =s_0}ds\, w(s) \Pi (s)\ ,
\label{fesrbasic}\end{equation}
and hence have an alternate, OPE representation for sufficiently large $s_0$. 
The FESR's corresponding to $R_{V/A;ij}^{(k,m)}$ are called
the ``$(k,m)$ spectral weight sum rules''. The purely $J=0$ term
in an ``inclusive'' sum rule (one with both $J=0$ and $0+1$
spectral contributions) is called ``longitudinal'' in what follows.

$V_{us}$ is extracted using flavor-breaking combinations
\begin{equation}
\delta R^w(s_0)\, =\, \left[ R^w_{ud}(s_0)/\vert V_{ud}\vert^2\right]
\, -\, \left[ R^w_{us}(s_0)/\vert V_{us}\vert^2\right]\ ,
\label{tauvusbasicidea}\end{equation}
whose OPE representation, $\delta R^w_{OPE}(s_0)$, begins 
at dimension $D=2$, with a term proportional to $m_s^2$. 
Experimental determinations of $R^w_{ud,us}(s_0)$ over a range of 
$s_0$ and $w(s)$ then allow $V_{us}$ (and in principle also $m_s$)
to be extracted~\cite{pichetalvus,kmcwvus,kmcwmsvus}. Explicitly, from
Eq.~(\ref{tauvusbasicidea}), one has~\cite{pichetalvus}
\begin{equation}
\vert V_{us}\vert \, =\, \sqrt{ R^w_{us}(s_0)/
\left[ {\frac{R^w_{ud}(s_0)}{\vert V_{ud}\vert^2}}
\, -\, \delta R^w_{OPE}(s_0)\right]}\ .
\label{tauvussolution}\end{equation}
The approach is favorable for the determination of $V_{us}$ because, 
at scales $\sim 2-3\ {\rm GeV}^2$, and for 
weights used in the literature, $\delta R^w_{OPE}(s_0)$ is
{\it much} smaller than the individual flavor $ud$ and $us$ spectral 
integrals (typically at the few-to-several-$\%$ level).
An uncertainty $\Delta \left(\delta R^w_{OPE}(s_0)\right)$
in $\delta R^w_{OPE}(s_0)$ thus produces a fractional uncertainty
$\left( \simeq \Delta \left(\delta R^w_{OPE}(s_0)\right)/2\, 
R^w_{ud}(s_0)\right)$
in $\vert V_{us}\vert$ {\it much} smaller than that on
$\delta R^w_{OPE}(s_0)$ itself. Moderate precision for
$\delta R^w_{OPE}(s_0)$ thus suffices for a high precision determination of
$\vert V_{us}\vert$, provided experimental errors can be brought under 
control~\cite{pichetalvus}. At present, the accuracy of the
method is limited by the sizeable (statistics limited) uncertainties on the 
$us$ spectral distribution (which produce $\sim 3-4\%$
uncertainties on the $us$ spectral integrals, and hence $1.5-2\%$
errors on $\vert V_{us}\vert$). 
These errors will be radically reduced in the near future
through analyses, already in progress, 
of the new $\tau$ decay data from BABAR and BELLE.
We thus concentrate here on the issue of 
controlling OPE-induced errors.

\section{Some Technical Complications}
The results reported below are all based on the ALEPH $ud$ and $us$
spectral data~\cite{alephdata}, ALEPH being the only collaboration 
which has made its $us$ covariance matrix publicly available. The 
reported results have been rescaled to take into account current
values of the various branching fractions~\cite{pdg06}. In the
case of the $us$ data, this is done following the prescription
described in Refs.~\cite{alephusrescaling}. Details of the
OPE input employed, and the method used to estimate the $D=2$ truncation 
uncertainty (which is much more conservative than those used previously
in the literature) may be found in Ref.~\cite{kmcwmsvus}.

The first important technical complication is the very bad behavior of the
integrated longitudinal $D=2$ OPE series. Not only does the series show no 
sign of converging, even at the maximum scale, $s_0=m_\tau^2$, allowed by
$\tau$ decay kinematics~\cite{bck05,longconv} but, even worse, 
it badly violates spectral positivity constraints for 
{\it all} truncation schemes used in the literature~\cite{longposviol}. 
Fortunately, the problem is easily handled phenomenologically. 
The basic reason is that the longitudinal 
decay contributions are, for a combination 
of chiral and kinematic reasons, strongly dominated by the $\pi$ and $K$ pole 
terms~\cite{longposviol,km00}. These terms are very accurately known. The
residual non-pole (``continuum'') contributions can, moreover,
be well-constrained using the approaches described in
Refs.~\cite{pichetalvus,km00,longposviol,mkps,jop}. Bin-by-bin
subtractions of the longitudinal contributions to the experimental
decay distributions can thus be made, allowing direct determinations
of the $(0+1)$ spectral functions, from which FESR's not afflicted
by the problematic longitudinal $D=2$ OPE behavior can be constructed.
{\it The extremely bad behavior of the longitudinal OPE series makes
inclusive analyses employing the longitudinal OPE representation
untenable, and the longitudinal subtraction process unavoidable.}

The second technical complication is the slow convergence
of the $D=2$ $J=0+1$ OPE series, which is known exactly to
$O(\alpha_s^3)$ and given by~\cite{bck05}
\begin{eqnarray}
&&\left[\Delta\Pi (Q^2)\right]^{OPE}_{D=2}\, =\, {\frac{3}{2\pi^2}}\,
{\frac{\bar{m}_s}{Q^2}} \left[ 1+2.333 \bar{a}+\right. \nonumber\\
&&\left. 19.933 \bar{a}^2 +208.746 \bar{a}^3+(2378\pm 200)\bar{a}^4
+\cdots\right]
\label{d2form}\end{eqnarray}
where $\bar{a}=\alpha_s(Q^2)/\pi$ and $\bar{m}_s=m_s(Q^2)$, 
with $\alpha_s(Q^2)$ and $m_s(Q^2)$ the running coupling and strange quark 
mass in the $\overline{MS}$ scheme. 
Since independent high-scale determinations of $\alpha_s(M_Z)$~\cite{pdg06}
correspond to $\bar{a}(m_\tau^2)\simeq 0.10-0.11$, the convergence
at the spacelike point on $\vert s\vert =s_0$, for $s_0\leq m_\tau^2$,
is marginal at best. For the $J=0+1$, $(k,0)$ spectral weights,
$w^{(k,0)}(y)=(1+2y)(1-y)^{k+2}$, with $y=s/s_0=e^{i\phi}$, 
the factor $\vert 1-y\vert^{k+2} = \vert 2\, sin(\phi /2)\vert^{k+2}$ 
peaks more and more strongly in the spacelike direction with increasing $k$.
Slow convergence of the integrated $D=2$ series, deteriorating with 
increasing $k$, is thus expected for the $(k,0)$ spectral weights, an
expectation borne out by the results of 
Refs.~\cite{bck05,kmcwvus,kmcwmsvus}. Fig.~1 of 
Ref.~\cite{kmcwmsvus} also shows no consistent common 
$m_s$, $V_{us}$ fit region exists for the $O(\bar{a}^4)$-truncated version
of those $(k,0)$ spectral weight sum rules employed previously in the 
literature.

Since terms higher order in $\alpha_s$ are relatively more important
at lower scales, a signal of the premature truncation of a slowly
converging OPE series is an unphysical $s_0$ dependence to the truncated
sum. Such an unphysical dependence can be exposed by comparing the
truncated OPE and corresponding experimental spectral integrals over a range 
of $s_0$. This comparison is shown for the $(0,0)$ spectral weight, for 
a range of input $m_s\equiv m_s(2\ {\rm GeV})$, in Fig.~1.
The value of $\vert V_{us}\vert$ needed for the spectral integral differences
is obtained by fitting at $s_0=m_\tau^2$.
It is clear that no $m_s$ exists giving 
a good match between the OPE and spectral integrals.
One should bear in mind that very strong correlations
exist amongst the OPE integrals for different $s_0$ and, similarly,
amongst the spectral integrals for different $s_0$.
The results thus clearly point to problems with controlling OPE
uncertainties in the $(k,0)$ spectral weight sum rules. That
there is no obvious way to reduce the problematic $D=2$ OPE truncation 
component of this uncertainty represents an important limitation to
the use of the $(0,0)$ spectral weight sum rule as a means of determining
$\vert V_{us}\vert$.

\begin{figure}
\unitlength1cm
\begin{minipage}[t]{8.6cm}
\begin{picture}(8.5,13.2)
\epsfig{figure=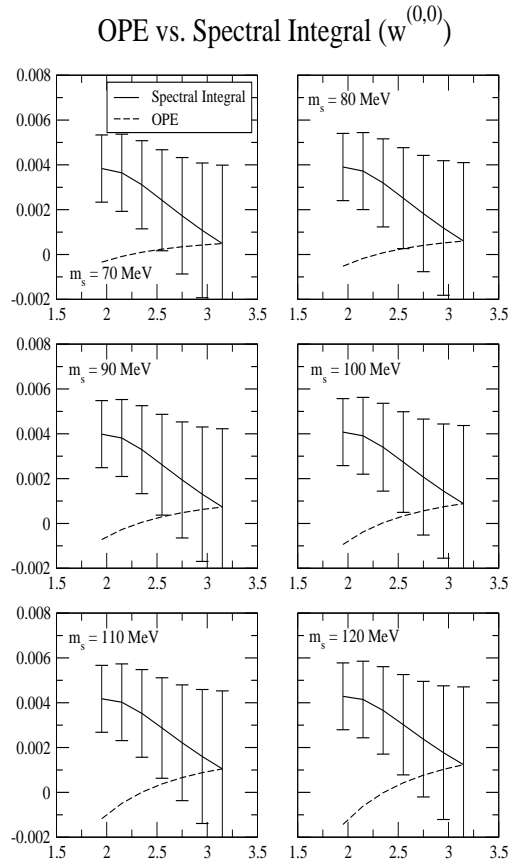,height=13.2cm,width=8.4cm}
\end{picture}\qquad\quad
\end{minipage}
\label{w00msrange}
\caption{$s_0$-stability plots for the $(0,0)$
spectral weight. The horizontal axis shows $s_0$ in GeV$^2$.} 
\end{figure}

\begin{figure*}
\unitlength1cm
\begin{minipage}[t]{8.6cm}
\begin{picture}(8.5,13.2)
\epsfig{figure=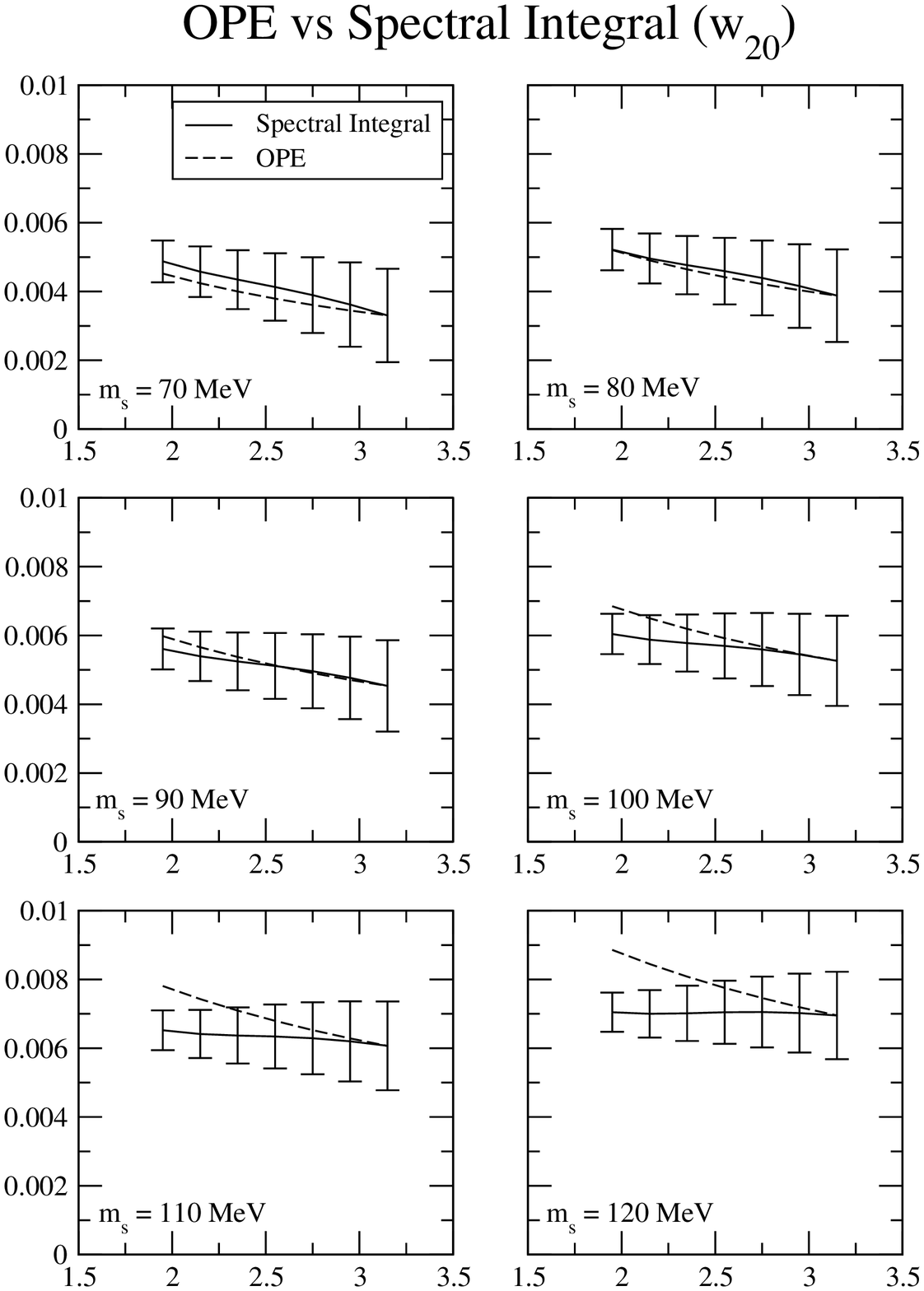,height=13.2cm,width=8.4cm}
\end{picture}
\end{minipage}
\hfill
\unitlength1cm
\begin{minipage}[t]{8.6cm}
\begin{picture}(8.5,13.2)
\epsfig{figure=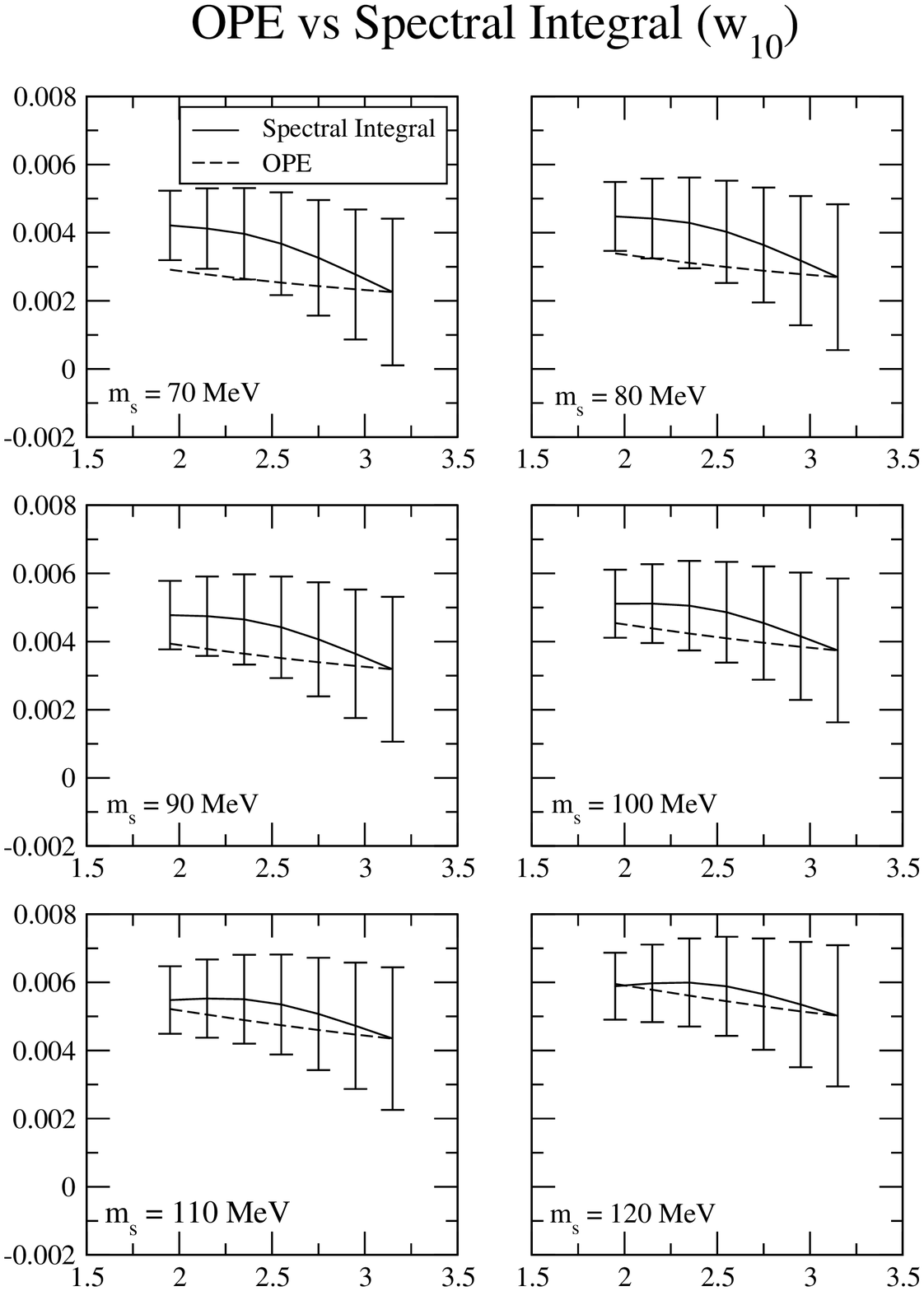,height=13.2cm,width=8.4cm}
\end{picture}
\end{minipage}
\caption{$s_0$-stability plots for $w_{20}$ (left panel) and
$w_{10}$ (right panel). The horizontal axis shows $s_0$ in GeV$^2$.}
\label{w20w10msrange}
\end{figure*}

The fact that $\vert \alpha_s(Q^2)\vert$ decreases as $Q^2=-s$ moves away from
the spacelike point along the circle $\vert s\vert =s_0$ allows 
one to improve the convergence of the integrated $J=0+1$, $D=2$ OPE 
series by working with weights which emphasize those parts 
of the contour away from the spacelike point. 
Three weights, $w_{10}$, $\hat{w}_{10}$, and $w_{20}$, designed specifically
to produce improved convergence (in addition to improved control of
other features of the OPE analysis) were introduced in Ref.~\cite{km00}.
A fourth such weight, $w_8$, was introduced in Ref.~\cite{kmcwmsvus}.
The improved convergence, and resulting much improved $s_0$-stability
of $\vert V_{us}\vert$, have been demonstrated
previously~\cite{km00,kmcwvus,kmcwmsvus}. Here, in 
Fig.~\ref{w20w10msrange}, we show the analogues of Fig.~1
for the weights $w_{20}$ and $w_{10}$ (the results for $\hat{w}_{10}$
and $w_8$ are intermediate in quality, and hence not shown explicitly).
The contrast to the $(0,0)$ spectral weight case is
immediately evident. We have also shown elsewhere (see Fig.~4 of
Ref.~\cite{kmcwmsvus}) that an excellent common fit region for 
$m_s$ and $\vert V_{us}\vert$ exists for the four non-spectral weights
considered here, in sharp contrast to the lack of consistency found for the
$(k,0)$ spectral weights.

\section{Results and Prospects}
Single non-spectral weight, $s_0=m_\tau^2$, fits, using as input 
the average, $m_s(2\ {\rm GeV})=94\pm 6$ MeV, of strange scalar and 
pseudoscalar sum rule~\cite{sumrulems} and $N_f=2+1$ lattice 
results~\cite{latticems}, 
advocated in the last of Refs.~\cite{jop}, yield
$\vert V_{us}\vert =0.2209\pm 0.0029_{exp}\pm 0.0017_{th}$ for $w_{20}$, 
$0.2210\pm 0.0030_{exp}\pm 0.0010_{th}$ for $\hat{w}_{10}$, 
$0.2206\pm 0.0032_{exp}\pm 0.0007_{th}$ for $w_{10}$, 
and $0.2218\pm 0.0037_{exp}\pm 0.0009_{th}$ for $w_{8}$.
The errors are dominated by the large uncertainties  
on the experimental $us$ distribution. 

In view of the intrinsically slow convergence of the $J=0+1$, $D=2$ OPE 
series, it is worth testing further the nominally improved 
convergence of the non-spectral weight sum rules. We do so by ignoring 
external information on $m_s$ and performing a combined, $s_0=m_\tau^2$ 
fit for $m_s$ and $\vert V_{us}\vert$, obtaining 
$m_s(2\ {\rm GeV})=96\pm 31$ MeV and $\vert V_{us}\vert =0.2208\pm 0.0052$. 
The result for $m_s$ is seen to be in excellent agreement with the average
of independent sum rule and unquenched lattice determinations, already quoted 
above. This agreement justifies 
a combined $s_0=m_\tau^2$ fit with the external average, 
$m_s(2\ {\rm GeV})=94\pm 6$ MeV, as input. The result,
\begin{equation}
\vert V_{us}\vert =0.2209\pm 0.0031\ ,
\label{vusmainresult}\end{equation}
is compatible, within mutual errors, with both the recent $K_{\ell 3}$ 
determination, $\vert V_{us}\vert = 0.2249\pm 0.0019$~\cite{antonelli06}
and the recent $\Gamma [K_{\mu 2}]/\Gamma [\pi_{\mu 2}]$ determination, 
$\vert V_{us}\vert = 0.2223^{+0.0026}_{0.0013}$~\cite{latticems}.
Eq.~(\ref{vusmainresult}) represents our best determination of 
$\vert V_{us}\vert$ from present data.

As for future prospects, re-running the combined non-spectral
weight analysis with an assumed uncertainty of $\pm 5$ MeV on 
$m_s(2\ {\rm GeV})$ and $us$ spectral errors reduced by a factor of $5$ 
(a reduction expected to be readily achievable from the B-factory analyses), 
produces a total combined error $<0.0010$ on $\vert V_{us}\vert$.
Further improvements to the theoretical component of this error are
almost certainly possible, once the currently rather large $us$ 
errors above the $K^*$ are reduced, making it practical to construct, test, 
and work with additional weights 
having less strong suppressions of contributions from this part of the 
spectrum than do those employed above.

We conclude by discussing two aspects of the current data situation
which make it likely that higher central $\vert V_{us}\vert$ values
will be obtained from future analyses. The first point concerns the $us$ data. 
At present, the branching fractions for observed strange decay modes 
go down only to the $\sim 3\times 10^{-4}$ level ($\sim 1\%$ of the 
total strange branching fraction)~\cite{pdg06}. Adding the
contribution of any additional, as-yet-undetected strange mode will
necessarily increase $\vert V_{us}\vert$. The increase from a mode 
with branching fraction $\sim 1\times 10^{-4}$, for example, 
would be $\sim 0.0003-0.0004$. With the B-factory experiments
already reporting preliminary results for strange mode branching fractions 
at the few-to-several $10^{-5}$ level~\cite{tauckm06}, 
it is clear that such missing mode contributions will no longer be 
a problem once the BABAR and BELLE analyses are completed. The second 
point concerns the $ud$ data. With quoted errors, current experimental $ud$ 
uncertainties produce an error of $\sim 1/4\, \%$ on 
$\vert V_{us}\vert$. One should, however, bear in mind that significant 
discrepancies exist between the version of $\rho^{(0+1)}_{V;ud}(s)$ 
measured in $\tau$ decay and that implied by CVC and 
electroproduction (EM) data, even after all known isospin-breaking corrections
have been taken into account~\cite{dehz06}. The discrepancies are much larger
than the nominally quoted errors on the $\tau$ decay data, the direct
$\tau$ measurements and EM expectations for the branching fractions of
the $\pi^-\pi^0\nu_\tau$ and $\pi^-\pi^-\pi^+\pi^0\nu_\tau$
modes differing by $4.5\sigma$ and $3.6\sigma$, respectively. 
Replacing the $\tau$ $\pi\pi$ data with the nominally equivalent EM results 
would raise $\vert V_{us}\vert$ by $\sim 0.0017-0.0018$. A similar
replacement of the $\pi^-\pi^-\pi^+\pi^0$ data would raise it
by a further $\sim 0.0016$. While at present the $\tau$ decay data 
is in better agreement
than the EM data with certain OPE constraints~\cite{kmgminus2},
there is definitely room in the $\tau$ decay data for somewhat
lower $ud$ branching fractions~\cite{belleprelim}, which would lead to larger
$\vert V_{us}\vert$. Resolving the $\tau$ versus EM discrepancy
is thus important for the determination of $\vert V_{us}\vert$
using hadronic $\tau$-decay-based sum rules, as well as for resolving the
status of the hadronic contribution to $(g-2)_\mu$ in the Standard
Model.

\begin{acknowledgments}
The financial support of the Natural Sciences and Engineering
Research Council of Canada is gratefully acknowledged.
\end{acknowledgments}

\end{document}